\documentclass[aps,prd, nofootinbib,twocolumn,showpacs,superscriptaddress]{revtex4}
\usepackage{amssymb,amsmath,graphicx,color,microtype}
\usepackage{graphicx}
\usepackage{txfonts}
\usepackage{epsf}
\usepackage{epstopdf}
\usepackage {amssymb}
\usepackage{multirow}
\newcommand{\nc}{\newcommand}
\nc{\ba}{\begin{eqnarray}}
\nc{\ea}{\end{eqnarray}}
\newcommand\be{\begin{equation}}
\newcommand\ee{\end{equation}}

\nc{\e}{{\bf{e}}}
\nc{\kk}{{\bf{k}}}
\nc{\pp}{{\bf{p}}}

\nc{\bfk}{{\bf{k}}}
\nc{\bfx}{{\bf{x}}}
\nc{\bfp}{{\bf{p}}}

\nc{\eH}{{\epsilon_H}}
\nc{\calP}{{\cal P}}
\nc{\im}{{ \mathrm{Im} } }

\begin{document}
\title{Detectability of Gravitational Wave from a population of Inspiralling Black Holes \\
 in Milky Way Mass Galaxies}

\author{Razieh Emami}
\email{razieh.emami_meibody@cfa.harvard.edu}
\affiliation{Center for Astrophysics, Harvard-Smithsonian, 60 Garden Street, Cambridge, MA 02138, USA}

\author{Abraham~Loeb}
\affiliation{Center for Astrophysics, Harvard-Smithsonian, 60 Garden Street, Cambridge, MA 02138, USA}

\begin{abstract}
We estimate the rate of inspiral for a population of stellar mass BHs in the star cluster around the super massive black hole at the center of Milky Way mass galaxies. 
Our approach is based on an orbit averaged Fokker Planck approach. This is then followed by a post-processing approach, which incorporates the impact of the angular momentum diffusion and the GW dissipation in the evolution of system. We make a sample of 10000 BHs with different initial semi-major and eccentricities with the  distribution of $f_c(a)/a$ and $e$, respectively. Where $f_c(a)$ refers to the phase-space distribution function for cth species. 
Angular momentum diffusion leads to an enhancement in the eccentricity of every system in the above sample  and so increases the rate of inspiral. We compute the  fraction of time that every system spends in the LISA band with the signal to noise ratio $\rm{SNR} \geq 8$.   Every system eventually approaches the loss-cone with a replenishment rate given by the diffusion rate of the cluster, $\mu/ \rm{Gyr}^{-1} \lesssim 1 $. 
This small rate reduces the total rate of the inspiral for individual MW mass galaxies with an estimate  $R_{obs} \lesssim 10^{-5} yr^{-1}$. It is expected though that a collection of  $N_{gal} \simeq 10^4$ MW mass galaxies lead to an observable GW signal in the LISA band. 

\end{abstract}

\maketitle

\section{Introduction}
The stellar cluster around the super massive black hole (SMBH), at the center of   Milky Way (MW) mass galaxies is made of stars in the main sequence as well as different BH species. The slow inspiral of BHs, or any other compact objects, around the SMBH may lead to the emission of gravitational waves (GW) which exhibits the environment around the SMBH in detail. This system is called extreme mass ratio inspiral (EMRI) with characteristic frequency as low as, $f \lesssim 10^{-2}$ Hz. Such a signal is therefore completely outside the accessibility of the ground based GW observatories, like LIGO. 

The proposed space born Laser Interferometer Space Antenna (LISA) is sensitive to the GW wavelengths above the earth size and is therefore suitable to detect GWs in much lower frequency range ($\sim 10^{-4} - 0.1 \rm{Hz}$).  LISA is expected to detect GWs from such a systems in different mass ranges for the SMBH as well as different redshifts \cite{Audley:2017drz, Gair:2010yu, Kocsis:2011dr, Yunes:2011ws, AmaroSeoane:2012je, Berry:2013ara, Babak:2017tow,Samsing:2018isx, DOrazio:2018jnv, Barack:2018yly,  Samsing:2018nxk, Hopman:2005vr,Preto:2009kd,Arca-Sedda:2017wea}. 

Here we simulate the star cluster around the SMBH at the center of MW mass galaxies and estimate the rate of the inspiral phase for  different BH species. Our simulations are based on an orbit averaged, 1D, Fokker Planck approach, implemented in the Phase Flow (PF) library inside publicly available code AGAMA  \footnote{https://github.com/GalacticDynamics-Oxford/Agama} \cite{Vasiliev:2017sbo}.  
Our setup contains stars in the main sequence, with $m_{\star} = 1 M_{\odot}$ and 4 different BH species. We use  a publicly available population synthesis package  called COSMIC \footnote{https://github.com/COSMIC-PopSynth/COSMIC} 
to infer the BH masses as a function of ZAMS mass, focusing only on one metallicity value $Z = 0.001$. In our setup, we have also considered the continuous star formation with a constant rate. We infer the initial conditions for the density profile of different species, the properties of the source and the initial mass of the SMBH using an MCMC  approach and by a direct comparison with the most recent observations of SgrA*.

We have neglected the resonant relaxation in our analysis,  including both of the scalar and vector resonances  \cite{Merritt:2015vxa, Merritt:2015xpa,Merritt:2015elb, Merritt:2015kba}.

Using a post-processing approach, we have also considered effects such as the angular momentum diffusion as well as the GW emission in the analysis. We make a sample of 10000 BHs in  a two-dimensional space $(a,e)$ and compute the fraction of the time that each of them spends in the LISA band. We compute the expected rate and the number of the events in LISA band from the MW mass galaxies after $T_{obs} = 10 $yrs. While the rate of inspiral for the individual BHs are rather low, our computations show that such a signal is expected to be seen in a collection of $N_{gal} = 10^4$ MW mass galaxies. 

The paper is organized as follows. Sec. \ref{setup} presents the simulation setup. Sec. \ref{initial-profile} obtains the initial profile of the stellar cluster. Sec. \ref{Dynamical-Track} shows the dynamical evolution of BHs including the angular momentum diffusion as well as the GW emission. Sec. \ref{GW} introduces the mode function of the GW. Sec. \ref{detect} computes the expected rate of inspiral for MW mass galaxies. We conclude in Sec. \ref{conc}. Additional technical details are given in Appendix \ref{Fokker-Planck} and \ref{fitting}.

\section{Simulating Stellar Cluster around SgrA*}
\label{setup}
We start with our simulation setup. As already mentioned above, in our simulation, we use the PF code inside the AGAMA  code and we consider stars in the main sequence plus 4 different BH species with masses being inferred from the COSMIC code. 
 Using the PF code, we simulate the time evolution of the phase space distribution function, $f_c$ , where sub-index $c$ refers to stars and BHs species  influenced by the central SMBH. 

\ba
\label{Fokker-Planck1}
\frac{\partial f_c (h,t)}{\partial t} = - \frac{\partial \textit{F}_c(h,t)}{\partial h} - \nu_c(h,t) f_{c}(h,t) + S(h, t),
\ea

There are few terms in Eq. (\ref{Fokker-Planck1}) that require further explanations. In Appendix \ref{Fokker-Planck}, we make some introductory remarks about each of them.

\section{Initial Density Profiles for stars and BHs}
\label{initial-profile}
Next  here we describe on the choice of initial conditions for the phase-space distribution function of different species as well as the initial mass of SMBH. These are very important parameters that affect the dynamics of the system significantly. Since we are dealing with coupled differential equations and as the system evolves for a significant amount of time, few \rm{Gyrs}, it is very challenging to infer these parameters. Owing to this complexity, in our analysis, we use an MCMC approach to infer them from a direct comparison with the most recent observations. We take the initial density profile to be the following spheroidal profile and we find its different parameters from our MCMC analysis,

\ba 
\label{spheroidal}
\rho(r) = \rho_0 \left(\frac{r}{R_{scale}}\right)^{-\gamma} \left[1 +\left( \frac{r}{R_{scale}} \right)^{\alpha}\right]^{(\gamma - \beta)/\alpha},
\ea

Eq. (\ref{spheroidal}) describes the initial density profile for stars and BHs and it contains 5 different parameters. We use this profile for both of stars and different BH species. One physical reason for this choice is that BHs are considered as the remnant of stars in the main sequence. It is therefore reasonable to expect that the shape of their initial density profiles are similar. As we explain in the following, the only exception would be in their overall mass/density 
normalization. In our analysis, we use the COSMIC code to find a map between the BH remnant mass vs the stellar mass. Throughout our analysis, we fix the metallicity at $Z = 0.001$. Work is in progress to generalize this approach to different metallicities. Using the remnant map we reconstruct the BH IMF and we compute the fractional normalization for every BH species. Table \ref{BH-Norm} presents the inferred BH mass as well as the fractional BH normalization for different BH species. Further details about this map can be found in Ref. \cite{Emami:2019mzi}. 
We note that the second column in Table \ref{BH-Norm} presents the ``fractional'' normalization in BH, $F[i]$. This means that the actual normalization in the density/Mass of BHs is given by $F[i]$ multiplied with the actual normalization of ZAMS, $M_{\star}$, which is a parameter in our fits. Finally, since BHs are the remnant of stars in the main sequence, we take the normalization of the stars to be $(F[0] = 1- \sum_i F[i]) \times M_{\star}$. 


\begin{table}[th!]
\centering
\caption{COSMIC inferred Map between BH mass and the fractional normalization in density/Mass compared with  ZAMS.}

\begin{tabular}{|lc|r} 
\hline
$\mathbf{Mass(M_{\odot})[i] }$& 
$\mathbf{ F[i]} $\\
\hline
8
& 6.95 $ \times 10^{-3}$
\\
\hline
16
&  3.11 $\times 10^{-2}$
\\
\hline
24
& 
5.01 $\times 10^{-2}$
\\
\hline
35 &
 7.62 $\times 10^{-3}$
 \\
\hline
\end{tabular}
\label{BH-Norm}
\end{table}

We have performed an MCMC analysis and by experience figured out that some of these parameters, like $\alpha$ and $\beta$, are not important. Owing to this fact, we fixed them at $\alpha = 4 $ and $\beta = 5$. So at the end, we are left with  3 free parameters in our density profile that must be fixed using the MCMC analysis. 
In addition, since the mass of SMBH is gradually increasing owing to the disrupted stars and swallowed BHs, its initial mass is another parameter that must be inferred from the direct MCMC analysis. This adds an extra parameter in the our fitting analysis. 

Finally, our source term, $S(h, t)$, also contains two parameters, the source fractional mass as well as the source radii. This adds two more parameters in the MCMC analysis. 
We split the total mass in different species in two different parts. The first part is the continuously formed mass in this species, $fraC_{source} F[\mu] M_{\star}$, with $fraC_{source}$ referring to the fraction of continuously born source and with $\mu = (0, 1, 2, 3,4) $ referring to stars and BH. The second part is the initial normalization $\left(1 - fraC_{source}\right) F[\mu] M_{\star}$.  

This yields a family of total 6 parameters which must be fitted using MCMC analysis. We present further details of the analysis in Appendix \ref{fitting} and in the following, we just present the final results for the above parameters. 
\ba 
\label{6-dim-10percent}
\gamma &=& 1.15_{-0.46}^{+ 0.39} ~~, \\
\log_{10}(R_\mathrm{scale} [pc]) &=& 1.31_{-0.70}^{+ 0.6} ~~, \\
\log_{10}(M_\mathrm{*,init}) &=& 7.64_{-0.06}^{+0.23} ~~ , \\
\log_{10}(M_\mathrm{\bullet,init})&=& 6.52_{-0.04}^{+0.02} ~~, \\
\mathrm{frac}_\mathrm{source} &=& 0.44_{-0.15}^{+0.08} ~~, \\
\log_{10}(r_\mathrm{source}) &=& 0.49_{-0.09}^{+0.08} ~~.
\ea

Hereafter, we use the best fit of above parameters in our PF simulations. We make a box of five different species, one for the stars in the main sequence,  plus 4 different BH species with the masses and fractional normalization given by Table \ref{BH-Norm}. 

\section{Tracking the dynamical evolution BHs around SMBH}
\label{Dynamical-Track}
So far we have been focusing on presenting our simulation setup as well as the selection of the initial conditions. Hereafter, we use these simulations and estimate the rate of inspiralling stellar mass BHs around the SMBH in MW mass galaxies. This is done in few steps. At first, we should sample the BHs around the SMBH, i.e. making some grids in the semi-major axes and the eccentricity values. As it turns out, the distribution function of the semi-major axes scales as $f_c(a,t)/a$ where $f_c(a,t)$ refers to the phase-space distribution which is computed from the PF code. We should emphasize here that the original output of the PF is in terms of the phase-space volume and so we need to make a transformation from the $h$ space to $a$ space. As for eccentricity, we choose it from a thermal distribution. Next, we put these values inside a set of differential equations for the semi-major axes and the eccentricity and make a post-processing step in tracking $(a, e)$ with time up until hey hit the loss-cone or up to a given maximum time (which is defined below). We  then compute the lifetime of each of these BHs as well as the fraction of the time that they spend inside the observable bands (as defined in the following). Finally, we shall combine this with the replenishment rate which is given by the diffusion timescale and can be computed from the PF code. The combination of all of these steps gives us a consistent expected rate for the inspiralling stellar mass BHs around the SMBH to be seen with different GW observatories such as LISA. 

In the following sub-sections, we go over all of these steps in detail and we ultimately estimate the expected rate of inspirall from the MW mass galaxies. 
\subsection{Samples of the initial conditions}
\label{initial-sample}
\begin{figure}[th!]
 \center
 \includegraphics[width=0.48\textwidth]{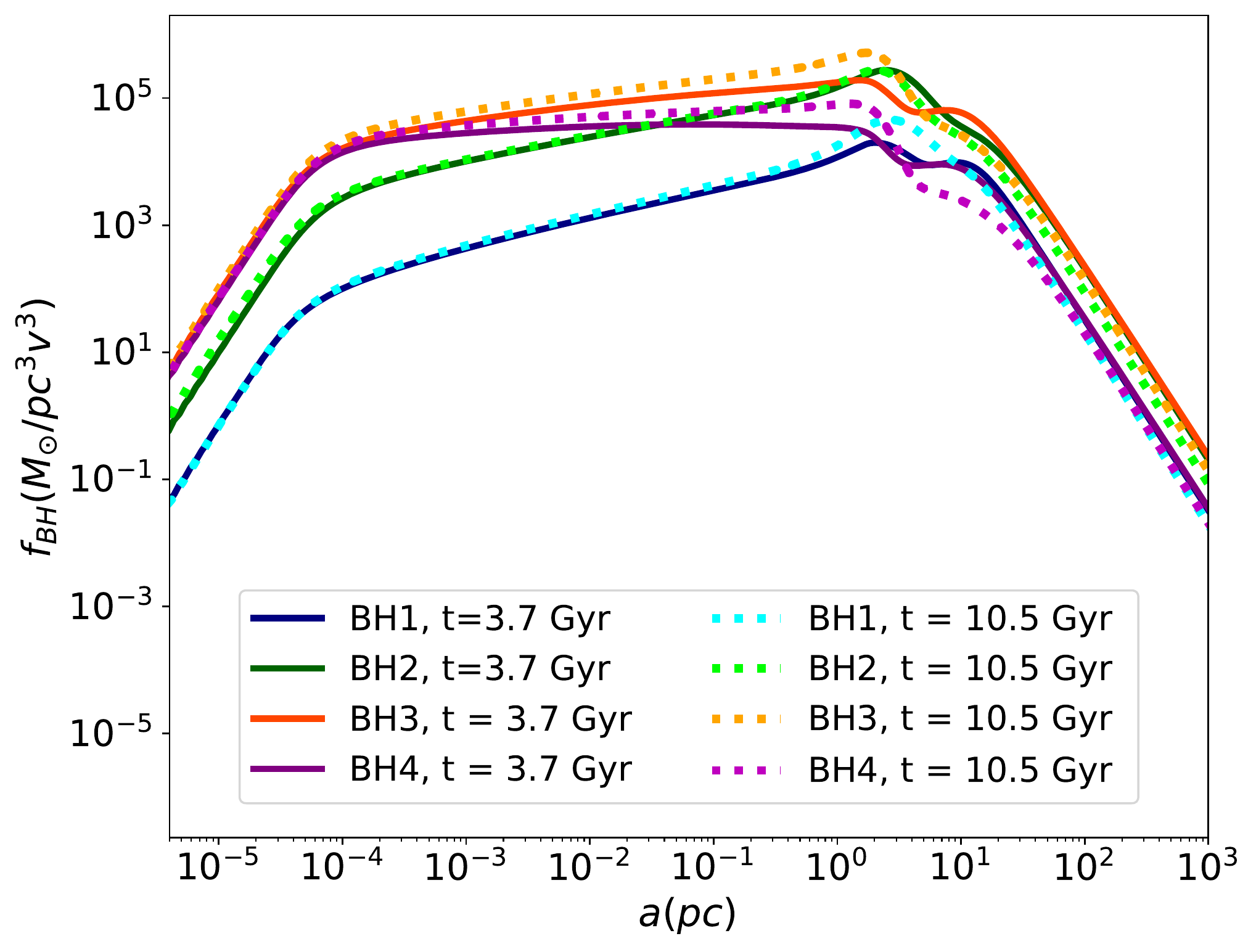}
\caption{ Behavior of $f(a, t)$ for different BHs. We present $f(a,t)$ for different BH species where different BHs with different masses are ordered as BH1, BH2, BH3 and BH4 , with an increasing mass from BH1 to BH4. }  
 \label{phse-space}
\end{figure}

\begin{figure*}[t!]
 \center
  \includegraphics[width=\textwidth]{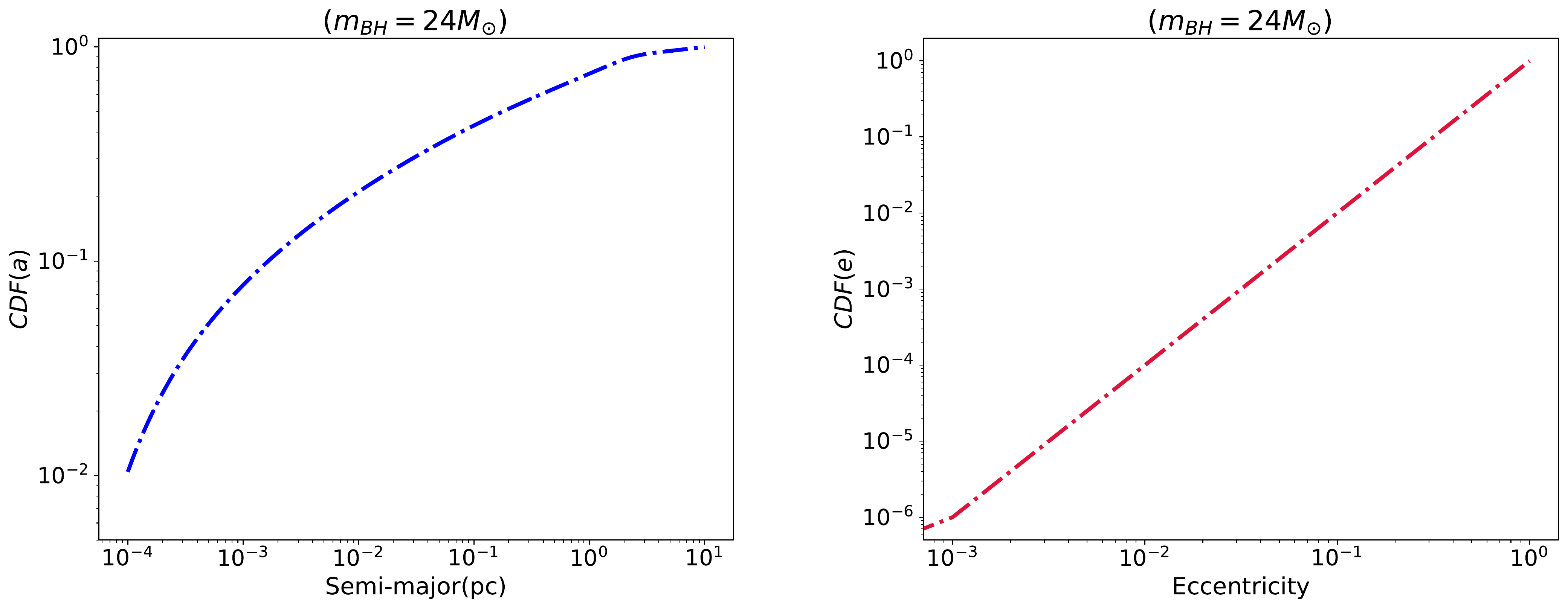}
\caption{ \textit{Left:} Cumulative distribution function of the semi-major axes, $a$, as chosen from $f(a)/a$ distribution for the BH  with  mass $m_{BH} = 24 M_{\odot}$.  \textit{Right:} CDF of the eccentricity taken from a thermal distribution. }  
 \label{CDF-a-e}
\end{figure*}

We start with an initial sample of BHs. We choose a total  $N_{tot} = 10^4$ pairs of initial values of $a$ and $e$, 100 values for each of them with the following distribution functions. Initial values of $a$ are selected from $f(a,t)/a$ distribution, where we have omitted the sub-index c for brevity. Since the phase-space distribution function shows some time-dependencies early on in the evolution, we allow the system to get relaxed for about $t = 3.7$ \rm{Gyrs} and with very little time evolution afterward.  Figure \ref{phse-space} presents the behavior of $f(a, t)$ for different BHs. The plot compares $f(a,t)$
of different BH species at two different times, namely at $t = 3.7$ \rm{Gyrs} as well as $t = 10.5$ \rm{Gyrs}. As it is clear from the plot, the star cluster get relaxed after about $t = 3.7$ \rm{Gyrs} and so we can take that value to represent a semi-static initial condition for the BHs. 
Here we specify different BHs as BH1, BH2, BH3 and BH4, where BH mass increases from BH1 to BH4. 
We take this relaxed distribution function and make a grid of semi-major axes in the range  $(7 \times 10^{-6} - 10)$ pc. Initial values of $e$, on the other hand, are adapted from a thermal distribution, $2e$. For each of these distribution functions we first compute the cumulative distribution function (CDF) and then we compute the inverse of CDF and make the samples. For example, Figure \ref{CDF-a-e} presents the CDF of $a$ (left panel) as well as $e$ (right panel) for the BH with mass $m_{BH} = 24 M_{\odot}$. 

Taking the above initial conditions, we evolve every systems for a maximum period of $t = 6.6 $\rm{Gyrs} or when they cross the loss-cone surface, which is determined in what follows. 
\subsection{Dynamical Evolution of semi-major axes and eccentricity}
Having presented the initial conditions of pairs of (a,e) in Sec. \ref{initial-sample}, we next study the dynamical evolution of every system. We include the impact of the GW as well as the angular momentum in our analysis. This leads to the following equations of motions, 

\ba 
\label{a-t-relation}
\frac{d}{dt}\left(a(t)\right) &=& -\frac{64}{5} \frac{G^3 m_{bh} M_{\bullet}M_{tot}}{c^5 a^3 (1-e^2)^{7/2}} \left(1 + \frac{73}{24} e^2 + \frac{37}{96} e^4 \right), \\
\label{e-t-relation}
\frac{d}{dt}\left(e^2(t)\right) &=& -\frac{608}{15} \frac{G^3 m_{bh} M_{\bullet}M_{tot} }{c^5 a^4 (1-e^2)^{5/2}} \left(1 + \frac{121}{304} e^2  \right) e^2  \nonumber\\
&&
+ \frac{\mu(a)}{\left(\alpha + \ln{\left(\frac{c^2 a}{16 G M_{\bullet}} \right) \left(1 - e^2 \right) } \right)}.
\ea
with $M_{tot} \equiv m_{bh} + M_{\bullet}$. Here $m_{BH}$ refers to the mass of every BHs while $M_{\bullet}$ refers to SMBH mass which is slightly growing in time, though in small range in our short lived BH sample. We then take it to be nearly the same as initial value from the PF code. The second term in Eq. (\ref{e-t-relation}) describes the angular momentum diffusion. This novel effect was ignored in the previous analysis and as it turns out, it is very important effect in enhancing the eccentricity dynamically. Although in completely different context, effectively this term acts similar to the famous Kozai-Lidov oscillations for triple systems. Owing to this, lots of our orbits do experience some horizontal evolution, along eccentricity axes, before they enter inside the GW bands. 
\begin{figure*}[t!]
 \center
  \includegraphics[width=\textwidth]{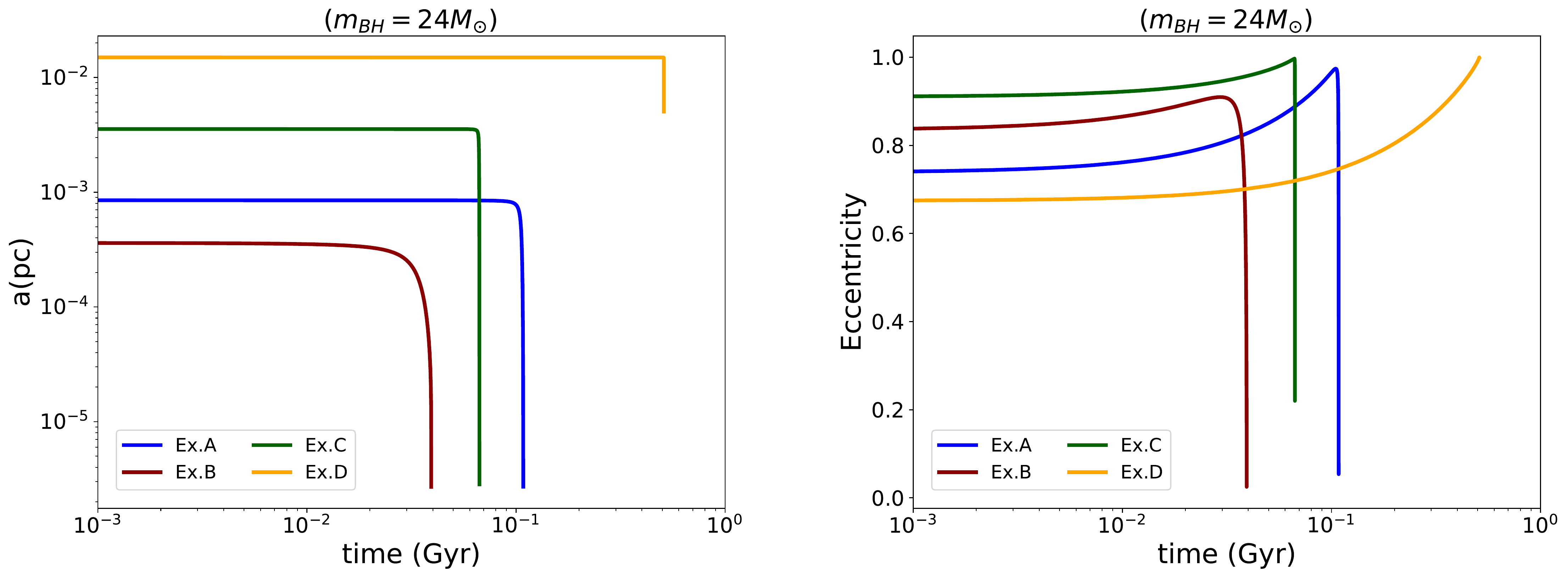}
\caption{ Time evolution of the semi-major axes, $a$, \textit{left}, and eccentricity, \textit{right}, for for few different initial conditions. Here we present the behavior of BH with mass $m_{BH} = 24 M_{\odot}$. }  
 \label{a-e-example}
\end{figure*} 

In Figure \ref{a-e-example}, we present the dynamical evolution of $a$ and $e$ for some of our initial conditions. Here we focus on the time-evolution for the case with $m_{BH} = 24 M_{\odot}$. Generally speaking, there are two main characteristic effects in the evolution of each systems that are worth mentioning. Angular momentum driven phase and the GW phase. In the former case, the system obeys an enhancement in the eccentricity while in the latter one, GW dominated regime, BHs get swallowed inward to the central BH and semi-major axes shrinks very rapidly.
In most cases we start with the first phase involving growth in eccentricity with no significant changes in the semi-major axes, and end up being in the second phase, where the system sinks very rapidly to the central BH with both of $a$ and $e$ evolve very quickly. This is indeed the case for examples A, B and C. However, in example D  we see a huge enhancement in the eccentricity until the system hits the boundary of the loss-cone and very rapidly gets capture by the central BH. In this case, the system does not have sufficient time to shrink the semi-major further and so $a$ does not change significantly throughout the evolution. 

\begin{figure}[th!]
 \centering
\includegraphics[width=0.45\textwidth]{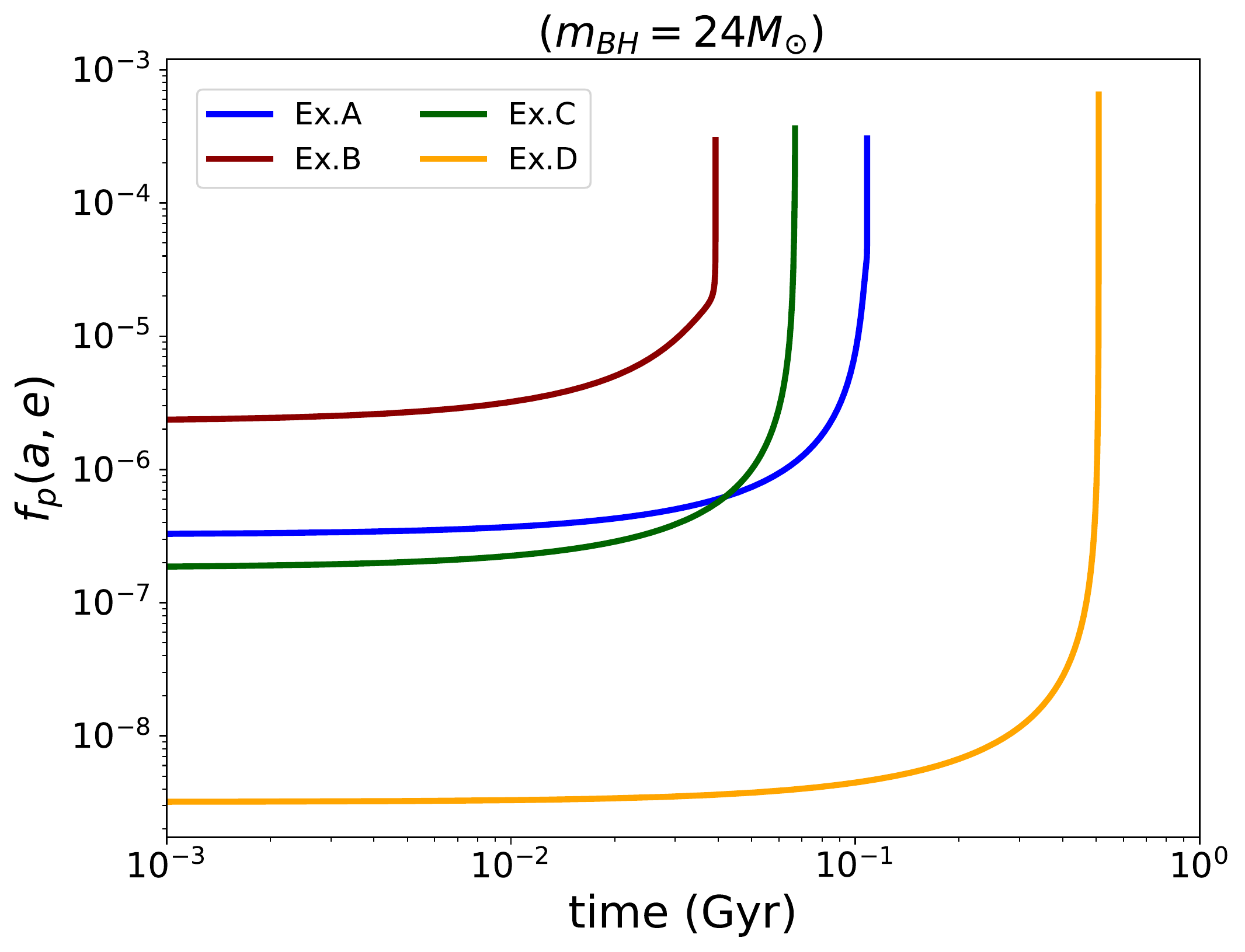}
\caption{The evolution of the peak frequency for some of the examples above. Here we present the behavior of BH with mass $m_{BH} = 24 M_{\odot}$. }  
 \label{peak-f}
\end{figure} 

In order to better understand the behavior of the above systems in the frequency plane, in Figure \ref{peak-f} we present the evolution of the eccentric orbital frequency, named as the peak frequency  \cite{Wen:2002km},

\ba 
\label{Frequency1}
f_p(e,a) = \frac{\sqrt{GM_{\bullet}}}{\pi} \left(\frac{\left(1 +e \right)^{1.1954}}{\left[a \left(1 - e^2\right)\right]^{3/2}}\right).
\ea
Being mostly in the angular momentum dominated phase, system D presents slightly different behavior with an extended time evolution compared with the rest of the systems. 

\section{GW Mode function and SNR}
\label{GW}
Using the above results for the dynamical evolution of pairs of $(a,e)$, here  we compute the mode function of the GW. Since we are dealing with the eccentric orbits, it is common to expand the mode function in terms of the harmonics  \cite{Wen:2002km}. In this basis, the characteristic mode function is defined as,

\ba
\label{GW-Mode-Function}
 h_{c,n} = \left( \frac{1}{ \pi d} \right) \sqrt{ \frac{2 \dot{E}_{n}}{\dot{f}_n}}.
\ea
where $n$ refers to the nth harmonic and $d = 7.8 $ \rm{kpc} describes the distance to the source. In addition, we have 

\ba 
\label{dotE}
\dot{E}_{n} &=& (32/5) \mathcal{M}^{10/3}_c  \left( 2 \pi f_{orb} \right)^{10/3}  g(n,e), \\
\label{forb}
f_{orb} &=& (1/2\pi)(G M_{\bullet}/a^3)^{1/2}, \\
\label{fn}
f_{n} &=& n f_{orb}, \\
\label{Chirp-Mass}
\mathcal{M}_c &\equiv & \frac{\left(m_{BH} M_{\bullet} \right) ^{3/5}}{ \left( m_{BH} + M_{\bullet} \right)^{1/5}}.
\ea

Furthermore, the signal to noise ratio (\rm{SNR}) is defined as, 
\ba 
\label{SNR1}
\left(\rm{SNR}\right)^2 = 2 \sum_{n = 1}^{n_{max}} \int \frac{h^2_{c,n}}{ f_n S(f_n)} d \ln{f_n},
\ea
with $n_{max} $ referring to the maximum number of the harmonics. Here we take this number 
 to be maximum $10^5$.  Also $S(f_n)$ refers to the LISA noise power. As we stop the integration at the loss-cone, we  directly checked that in all of the cases we consider we get $\dot{f}/f \ll T_{obs} \simeq 10$yrs. As a results, we can expand the integral in Eq. (\ref{SNR1}) as,

\ba 
\label{SNR2}
\left(\rm{SNR}\right)^2 = \left(\frac{512}{5 d^2}\right) T_{obs} \frac{\left(G M_{c} \right)^{10/3}}{c^8} \left( 2 \pi f_{orb} \right)^{4/3} \sum_{n = 1}^{n_{max}} \frac{g_n(e)}{n^2 S(f_n)}. \nonumber\\
\ea
Hereafter we choose the criteria of having $\rm{SNR} = 8 $ in inferring the detectable number of systems.

\section{Detectability of the GW signal from the MW Mass Galaxies}
\label{detect}

Having presented the formalism in computing the \rm{SNR}, here we consider the detectability of the GW signal from inspiralling BHs in MW mass galaxies, with an especial focus on SgrA*. For this purpose, we first invent a generic formalism in computing the rate of events from SgrA*. Then, we compute the expected number of signals from inspiralling BHs around SgrA*. Using our generic formalism, we estimate the inspiral rate and number for LISA.

\subsection{Estimation of inspiral rate}
\label{Rate-Estimation}

Next we describe our method in estimating the total rate of expected inspiral for MW mass galaxies. 

We start with the initial condition for every pairs of $(a,e)$,  as described above, and we evolve them with time either to the time of loss-cone crossing , which is defined as $a(1-e) = 8 G M_{\bullet}$, or up to maximum time $t = 6.6$ \rm{Gyrs}. Then we use Eq. (\ref{SNR2}) and  compute \rm{SNR} for every system during the entire of their ``lifetime" . Here the ``lifetime" is defined from $t = 0 $ to minimum between their loss-cone crossing time and $t = 6.6$ \rm{Gyrs}.  For every system, we calculate the total duration of time that it  has $SNR \geq 8$ with a frequency in the specific observation band, as describe below. We 
divide this duration, named as ``observable period", to the total lifetime of the system.

As an  example, suppose that we have a system with  total lifetime of $t = 10^{-6}$ \rm{Gyrs} and it spends a period of $\Delta t = 10^{-7}$ \rm{Gyrs} in the LISA band,. The weighting factor for such a system is given by $w = 10^{-7}/10^{-6}$.

\begin{figure}[th!]
 \centering
\includegraphics[width=0.45\textwidth]{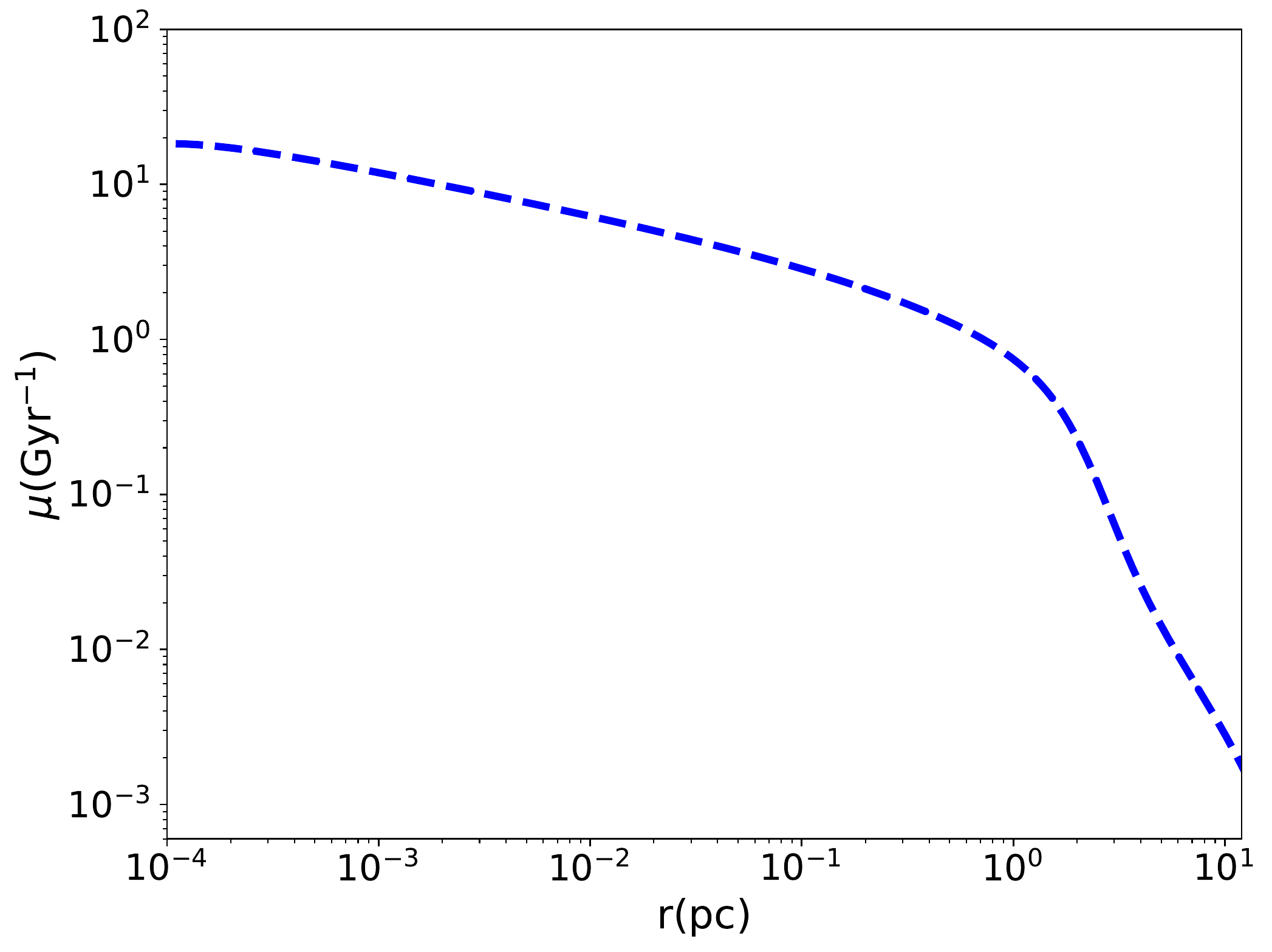}
\caption{Diffusion rate in a MW mass star cluster.  }  
 \label{diffusion}
\end{figure}

The above fraction must be multiplied with the replenishment factor, which is basically the rate of replacing every evolved system with the newer system. This is driven from the Phase-Flow code and is set with the rate of diffusion as given by $\mu$. Figure \ref{diffusion} presents the rate of the diffusion for a MW mass galaxy. As is seen from the plot the rate is quite small and so the replenishment rate is expected to be very small. Owing to this fact the expected inspiral rate for a single MW mass galaxy is low.

Combining the above factors, we get the following combined rate factor (hereafter $w(i)$)  for the ith system as, 

\ba 
\label{ith-Weight}
W(i) = \left(\frac{\Delta t_i}{t_i} \right) \times \mu_i,
\ea
where $\Delta t_i$, $t_i$ and $\mu_i$ refer to the observable period,  total lifetime and the diffusion rate for ith system, respectively. 

Next, we shall re-scale the above number to the actual number of BHs in the galactic center interior to the radius $r = r_{sam} = 10 \rm{pc}$, as the  upper limit in our sample.  This is done by the factor $N_{BH}(r \leq r_{sam})/N_{tot}$ where $N_{BH}(r \leq r_{sam}) \equiv M_{BH}(r \leq r_{sam})/m_{BH}$ describes the number of BHs with the given mass $m_{BH}$ interior to the radii $r = r_{sam} = 10 pc$ and $N_{tot} = 10000$ is the total number of the samples in our post-processing. Therefore, the final re-scaled rate for ith system would be, hereafter inferred as $W_{tot}(i)$,

\begin{table}[th!]
\centering
\caption{Critical semi-major axes above which the loss-cone timescale is shorter than the GW timescale. Therefore BHs do not emit GWs and scatter inward or outward to larger radii.}
\label{a-crit_tab11}
\begin{tabular}{|lcc|r} 
		\hline
		 $\mathbf{ m_{BH} (M_{\odot})}$ & ~~~
        $\mathbf{ a_{crit}(pc)}$ &
        \\
		\hline
       $8$ & 
        $0.01$ &
        \\
        \hline
        $16$ & 
        $~ 0.029$ & 
        \\
        \hline
        $24$ & 
        $0.04$ &
        \\
        \hline
        $35$ & 
        $0.054$  &
        \\
        \hline
    	\end{tabular}
\end{table}

\begin{figure*}[t!]
\center
\includegraphics[width=1.\textwidth]{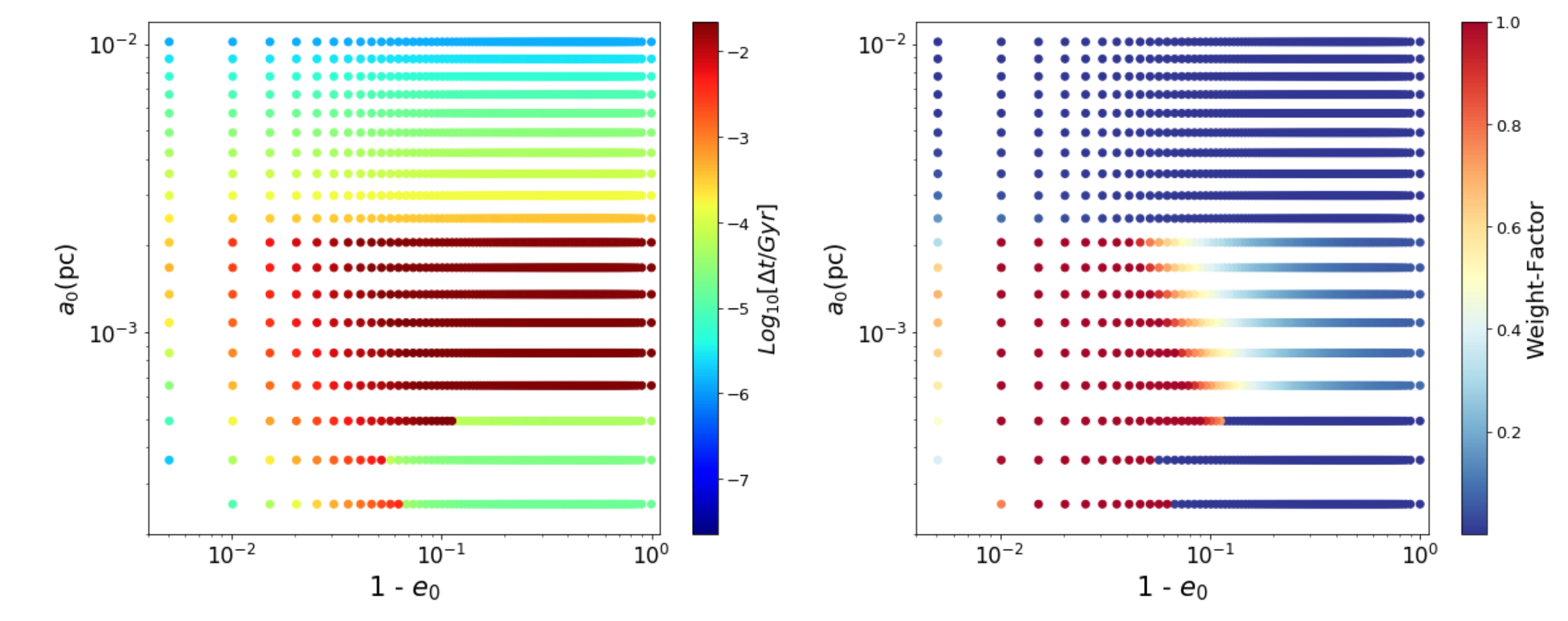}
\caption{Observable time (left) and the fraction of time (right) in the LISA band for one of the BHs in the system with mass $m_{BH} = 24 M_{\odot}$. }
\label{fraction-life}
\end{figure*}

\ba 
\label{Total-Weight}
W_{tot}(i) = \left(\frac{N_{BH}(r \leq r_{sam})}{N_{tot}} \right) \times W(i),
\ea
In addition, we should also eliminate cases with semi-major axes above a critical value, hereafter $a_{crit}$, for which the timescale of GW, $\tau_{GW} \equiv |a/\dot{a}|$, is longer than the associated loss-cone timescale $t_{LC} \equiv \left(L_{LC}/L_{cir} \right)^2 t_{Rel}$. Here the loss-cone timescale is defined as the timescale associated with a change in angular momentum by the order of  loss-cone \cite{Hopman:2006xn}. 
Combining with the loss-cone surface, $a(1-e) = 8 G M_{\bullet}$, we can estimate the critical semi-major axes above which $\tau_{GW} > t_{LC}$. As specified in \cite{Hopman:2006xn}, systems with $a>a_{crit}$ scatter either inward the loss-cone or outward to some larger orbits without emitting any GWs. Therefore they are not source of the inspiral phase and we should  remove them from our samples. Table \ref{a-crit_tab11} estimates $a_{crit}$ for different BH masses in our system. Interestingly the value of $a_{crit}$ increases for larger BH masses. That makes sense since the characteristic timescale associated with the GW get suppressed for heavier BHs though the loss-cone timescale remains the same. Therefore we may get further out from the center and yet be in the GW regime. 

In order to get some intuitions about the above abstract definitions, in Fig. \ref{fraction-life} we present the color-plot of  logarithm of the observable time, $ \log_{10} \Delta t/\rm{Gyr}$, (left) as well as the weighting factor, as the fraction of lifetime in the LISA band or $\Delta t / t$, (right) in the plane of initial semi-major and eccentricity $(a_0, e_0)$. 

From the plot we may infer few interesting regimes in the $(a_0, e_0)$ plane: 
starting from larger values of $a_0$ there is obviously less chance for the system to be in the LISA band. This is shown up in both of these plots with small value of $\Delta t$ and the weighting factor. Lower values of $a_0$, show more interesting behavior in a manner that depends on the initial eccentricity as well. Larger $e_0$ spends more in the LISA band. Decreasing $e_0$ suppresses the lifetime in LISA as well as the weighting factor. There is however some intermediate stage where dynamical evolution becomes non-trivial due to the angular momentum diffusion and for them we see that in some cases we may get slightly larger weighting factor as the enhancement in the eccentricity is larger. This is a dynamical effects, due to further enhancement of $e$. This can be seen from Figure \ref{a-e-example} in which Ex. A shows more enhancement in the eccentricity though its initial semi-major axes is larger than Ex. B. This shows through a  slight enhancement of the weighting factor in a very small interval of $(a_0, e_0)$ in the right panel of Figure \ref{fraction-life}.  Finally, for very small values of $a_0$ the system spends its entire evolution in the LISA band. Since these systems are rare, we do not show them in the figure.

The total rate of inspiral, referred as $R_{obs}$, is then,
\ba 
\label{Observable-Fraction}
R_{obs} &\equiv& \sum_i W_{tot}(i) \nonumber\\
&=& \sum_i  \left(\frac{N_{BH}(r \leq r_{sam})}{N_{tot}} \right)  \left(\frac{\Delta t_i}{t_i} \right) \mu_i.
\ea

yielding the total number of inspiralling BHs  after $T_{obs}  $,

\ba 
\label{tot-Number1}
N_{lisa} = R_{obs} \times T_{obs}.
\ea
 Eqs. (\ref {Observable-Fraction}) and (\ref{tot-Number1}) are the key results of this paper. 
In the following, we use these two equations and estimate the inspiral rate and number of expected signals from MW mass galaxies with LISA. We compute the rate and number for different BH masses in our sample and different observational time, relevant for the expected number of signals. 

Table \ref{number} presents the expected rate as well as the number per one MW mass galaxies. It can be seen from the table that the expected rate is rather small for a typical MW mass galaxies but it can be seen in a sample of about $N_{gal} \simeq 10^4$ MW mass galaxies. 

\begin{table}
\centering
\caption{Expected rate of inspiral and the number of LISA sources per one Milky Way like galaxy.}
\label{number}
\begin{tabular}{|lcc|r} 
		\hline
		 $ m_{BH} (M_{\odot})$ & 
        $ R_{obs} (yr^{-1})$ &
        $ N_{lisa} (T_{obs} = 10 yr)$
        \\
		\hline
       $8$ & 
        $2 \times 10^{-6}$ &
        $2 \times 10^{-5}$ 
        \\
        \hline
        $16$ & 
        $1.5 \times 10^{-5} $ & 
        $1.5 \times 10^{-4} $
        \\
        \hline
        $24$ & 
        $3 \times 10^{-5}$ &
        $3 \times 10^{-4} $
        \\
        \hline
        $35$ & 
        $6 \times 10^{-6}$ &
         $6 \times 10^{-5}$
        \\
        \hline
    	\end{tabular}
\end{table}

\section{Conclusion}
\label{conc}
Using a hybrid tool made of an orbit averaged  Phase Flow code, as a library in the AGAMA code, as well as a binary stellar evolution code, COSMIC, we simulated the stellar cluster inside Milky Way like galaxies with a SMBH at the center. Our set up was made of stars in the main sequence, with $m_{\star} = 1 M_{\odot}$ and 4 different BH species with masses inferred from the initial metallicity, which is fixed in our setup at $Z = 0.001$. Using an MCMC approach and by a direct comparison with the most recent observations, we inferred the initial conditions for the density profile of the system. We used this setup and simulated the stellar cluster dynamically with time. Based on a post-processing approach, we computed the dynamical evolution of the semi-major axes as well as the eccentricity taking into account the impact of the angular momentum diffusion in the evolution. 
Interestingly, the angular momentum diffusion leads to an enhancement in the eccentricity of the system. Though in completely different setup, this is reminiscence of Kozai-Lidov oscillations in the context of triple systems \cite{Smadar}. 

We studied the detectability of the GW in this context from the inspiralling BHs around the SMBH. We constructed a sample of $10^4$ pairs of initial semi-major and eccentricity and evolved all of them with time and computed the SNR for all of them during the entire of their lifetime and inferred the fraction of time that every systems spend in the LISA band. We then weighted this number with the diffusion rate taken as a replenishment factor. Furthermore, we rescaled this number with the total number of BHs inside the $r = 10 \rm{pc}$ which is the upper limit in our samples. We also  eliminated  systems with $a>a_{crit}$ with $a_{crit}$ referring to a maximum semi-major axes above which  characteristic timescale of  GW is larger than the associated loss-cone time-scale. In this region, BHs may scatter either off the central BH to some larger orbits or just get swallowed to the center without emitting any GWs signal. Owing to this reason, we removed such systems from our samples entirely. 
The total rate and expected number of the event with LISA from the MW mass galaxies are presented in Eqs. (\ref{Observable-Fraction}) and (\ref{tot-Number1}), respectively. We presented the final rate and the expected number of the inspiralling BHs in Table \ref{number}. Owing to the small replenishment rate, the inspiral rate is rather low. The signal can nevertheless be seen if we consider a collection of about $ 10^4$ MW mass galaxies. Interestingly, the signal for individual BHs peaks at the peak of their initial normalization factor which is itself the peak of the BH mass-function \cite{Emami:2019mzi}. 

\textbf{Acknowledgment.} We thank Sownak Bose, John Forbes and Dan D'Orazio for the  helpful discussions. R.E. acknowledges the support by the Institute for Theory and Computation at the Center for Astrophysics. We also thank two anonymous referees for their insightful comments. 
This work was also supported in part by the Black Hole Initiative at Harvard University which is funded by a JTF grant. We thank the supercomputer facility at Harvard where most of the simulation work was done. 

\appendix

\section{Fokker-Planck approach}
\label{Fokker-Planck}
Here we introduce different terms enter in Eq. (\ref{Fokker-Planck1}). First of all, 
PF code used the phase space volume $h(E)$,  instead of energy, as the main variable in the analysis. 
$h(E)$ refers to an enclosed volume by the energy hypersurface $E$ \cite{Vasiliev:2017sbo}. This is defined by,
\ba 
\label{phase-space-volume}
h(E) = 16 \pi^2 \int_{\Phi(0)}^{E} dE' \int_{0}^{r_{max}(E')} r^2 \sqrt{2\left(E' - \Phi(r) \right)} dr,
\ea

with  $\Phi(r)$ describing the total gravitational potential of the system, 
\ba 
\label{Phi}
\Phi(r) &=& - \frac{G M_{\bullet}}{r} - 4 \pi G \sum_{i} \bigg{[} \left( \frac{1}{r}\right) \int_{0}^{r} dr' r'^2 \rho_i(r')  
\nonumber\\
&&
~~~~+ \int_{r}^{\infty}  dr' r'  \rho_i(r') \bigg{]}.
\ea
Furthermore $\rho_i(r)$ refer to mass density of every species. 

Another important parameter entering in Eq. (\ref{Fokker-Planck1}) is the mass flux through $h$, 
$\textit{F}_c(h,t)$, which is given by, 
\ba 
\label{mass-flux1}
\textit{F}_c(h,t) = A_c f_c + D \frac{\partial f_c}{\partial h},
\ea
with $A_c(h)$, $D(h)$ referring to advection and diffusion coefficient which are given by, 
\ba 
\label{advection1}
A_c(h)  &=& 16 \pi^2 G^2 \ln{\Lambda} ~ m_c \sum_i \int_{0}^{h} f_i(h') dh', \\
\label{diffusion1}
D(h) &=& 16 \pi^2 G^2 \ln{\Lambda}~ g(h) \sum_i  m_i \bigg{(} \int_{0}^{h} \frac{f_i(h') h'}{g(h')} dh' + \nonumber\\
&& ~~~
h  \int_{h}^{\infty} \frac{f_i(h')}{g(h')} dh' \bigg{)}.
\ea
We summed over all of different species in Eqs. (\ref{advection1}) and (\ref{diffusion1}). We choose a Coulomb logarithm \cite{Binney-Tremaine} $\ln{\Lambda} \simeq 10$ in our analysis. 

Next we introduce the sink term, $\nu_c(h,t)$. This terms is due to the consumption of stars and BHs nearby the central BH. For the stars in the main sequence it happens when they cross the tidal disruption radii, $R_{tid} \equiv R_{\star} \left(M_{\bullet}/M_{\star}\right)^{1/3} $ (with $M_{\star}$ and $R_{\star}$ refering to the mass and radius of a typical star). It is important to mention that not the whole mass of disrupted stars would be added to the central BH mass. Throughout our analysis we take this fraction to be $f_{dis} = 10 \%$. 
BHs, on the other hand, are getting swallowed by the SMBH if they cross the capture radius of the central BH, in a distance  $R_{cap} = 8 G M_{\bullet}/c^2$. The thought is that the mass of BHs are entirely added to the SMBHs. 
Following the approach of Ref. \cite{cohn1987}, the loss term is written as, 

\ba 
\label{loss-rate}
\nu = \frac{\mu(E)}{\alpha + \ln{\left(1/{\mathcal{R}}_{LC}\right)}} ~~~,~~~ \alpha \simeq \left(q ^2 + q^4 \right)^{1/4},
\ea
where $r_{LC}$ refers to the loss-cone radius, for stars and BHs, and hereafter we use $R_{tid}$ and $R_{cap}$ to infer this parameter, respectively. Finally, $q$  and $\mu(E)$ refer to  loss-cone filling factor  and orbit averaged diffusion coefficient, respectively. These parameters are presented  in details in \cite{Vasiliev:2017sbo}, and we encourage the interested readers to take a look at this reference for more details. 

Finally, $S(t,h)$ refers to the source term and refers to the continuous star formation in our system. 
Throughout this work, we choose this function constant with time and interior to a given radii $r_{source}$. More explicitly we take the source density profile to be proportional to $\sqrt{1/r - 1/r_{source}}$. So our source term has two different parameters, the mass fraction of the source and the source radii. As we mention in the following, we read these parameters from an MCMC fitting of the model with the observations.

\section{Initial conditions: An MCMC approach}
\label{fitting}
As already mentioned in the main body of paper, we are left with a family of 6 parameters that must be determined using a direct comparison with the recent observations. In our fitting analysis, we consult with the observational data from \cite{Schodel:2017vjf}. More specifically we use the left panel in Figure 9 of this paper. This gives us the flux density at different radial locations. Furthermore, we use the enclosed mass at two different radii, at 

\ba 
\label{condition}
M(r \leq 1pc) &=& 10^{6} M_{\odot}, \nonumber\\
M(r \leq 4pc) &=& 10^{7} M_{\odot}.
\ea
We also take the final mass of SMBH to be $M_{\bullet} = 4 \times 10^6 M_{\odot}$ after $t = 10.5$ \rm{Gyrs}. 

\begin{figure*}[t!]
\center
\includegraphics[width=1.\textwidth]{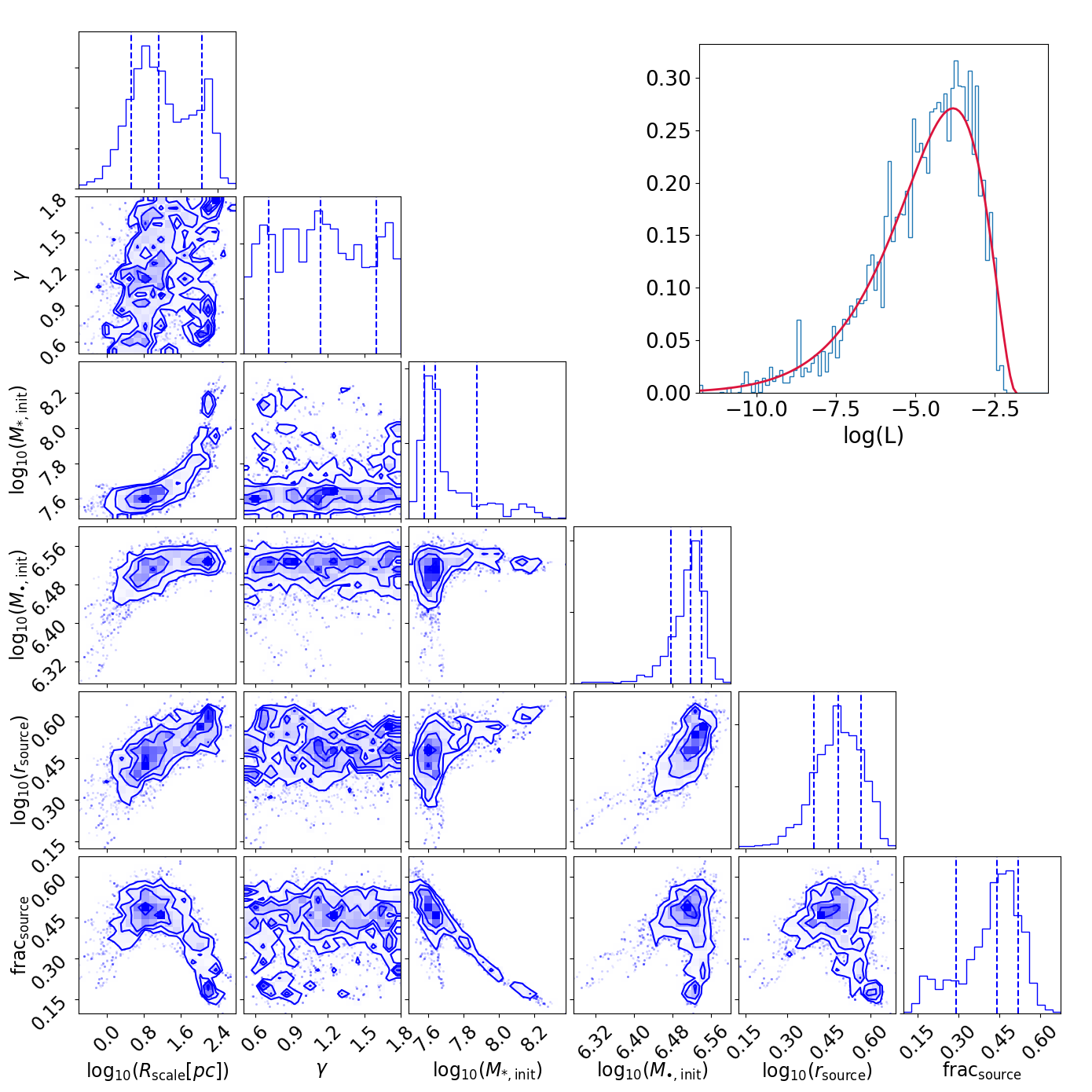}
\caption{\textbf{Posterior values of 6 parameter family inferred from MCMC analysis. }}
\label{10percent-tidal}
\end{figure*}

Using the above data, we find the proper values for above 6 dimensional parameter space including the
density slope, $\gamma$, and the logarithm of scaling radius, $\log_{10}(R_\mathrm{scale})$,  the logarithm of  initial total mass normalization, $\log_{10}(M_\mathrm{*,init})$, the logarithm of initial central BH mass, $\log_{10}(M_\mathrm{\bullet,init})$, fraction of the source mass, $\mathrm{frac}_\mathrm{source}$, and finally the logarithm of  source radius, $\log_{10}(r_\mathrm{source})$.

Figure \ref{10percent-tidal} presents the posterior plot for different parameters. Throughout our analysis, we use the best fit values of these parameters.


\end{document}